\documentclass[pra,aps,showpacs,twocolumn,amsmath,amssymb]{revtex4}

\usepackage{graphicx,dcolumn,bm}
\usepackage{psfrag}
\usepackage{pstricks}
\usepackage{color,rotating}

\definecolor{bleuMath}{rgb}{0.2472,0.24,0.6}
\definecolor{rougeMath}{rgb}{0.6,0.24,0.442893}
\definecolor{vertclair} {cmyk}{1,0,1,0.2}

\begin{document}

\title{Isotope shifts and hyperfine structure of the laser cooling Fe~I 358-nm line}

\author{N. Huet}
\author{M. Pettens}
\author{T. Bastin}
\affiliation{Institut de Physique Nucl\'eaire, Atomique et de Spectroscopie, Universit\'e de Li\`ege, 4000 Li\`ege, Belgium}

\date{\today}

\begin{abstract}
We report on the measurement of the isotope shifts of the $3d^74s \,\, a \, {}^5\!F_5 - 3d^74p \,\, z \, {}^5\!G^o_6$ Fe~I line at 358~nm between all four stable isotopes ${}^{54}$Fe, ${}^{56}$Fe, ${}^{57}$Fe and ${}^{58}$Fe, as well as the hyperfine structure of that line for ${}^{57}$Fe, the only stable isotope having a nonzero nuclear spin. This line is of primary importance for laser cooling applications.
In addition, an experimental value of the field and specific mass shift coefficients of the transition is reported as well as the hyperfine structure magnetic dipole coupling constant $A$ of the transition excited state in $^{57}$Fe, namely $A(3d^74p \,\, z \, {}^5\!G^o_6)=31.241(48)$~MHz. The measurements were carried out by means of laser-induced fluorescence spectroscopy performed on an isotope-enriched iron atomic beam. All measured frequency shifts are reported with uncertainties below the third percent level.
\end{abstract}


\maketitle

\section{Introduction}

In the past decades, laser cooling of atoms has given rise to fascinating new fields in atomic physics and has defined very active research lines in many related fields (see for instance the latest annual review~\cite{annualreviews}). If several atomic species can be nowadays laser cooled using well established laser schemes, still most have never been manipulated in this way, essentially because of a lack of suitable and affordable laser radiation. In this framework, an innovative cooling scheme for iron atoms has been recently proposed on the basis of 2 ultraviolet transitions at 372 and 358~nm, both accessible with commercial laser systems~\cite{iodinepaper}. Iron atom is a quite challenging atomic species for laser cooling. One of the main reasons comes from the lack of a suitable cooling transition from the ground state. A good transition for that purpose is rather provided by the $3d^74s \,\, a \, {}^5\!F_5 - 3d^74p \,\, z \, {}^5\!G^o_6$ transition at 358~nm~\cite{Nave94}, whose lower state lies at 6928~cm$^{-1}$ above ground state ($3d^64s^2 \,\, a \, {}^5\!D_4$)~\cite{Nave94,NIST} and is metastable with a lifetime of the order of several hundred seconds~\cite{Gre71}. The atoms can be optically pumped to this lower state from the Fe~I ground state through the 372-nm $3d^64s^2 \,\, a \, {}^5\!D_4 - 3d^64s4p \,\, z \, {}^5\!F^o_5$ resonance line~\cite{Nave94,NIST}. The transition $3d^74s \,\, a \, {}^5\!F_5 - 3d^64s4p \,\, z \, {}^5\!F^o_5$ at 501~nm can then serve as a good decay channel to finalize the optical pumping from the ground state.

Iron has four stable isotopes $^{54}$Fe, $^{56}$Fe, $^{57}$Fe and $^{58}$Fe, where $^{57}$Fe is the only one to have a nonzero nuclear spin $I = 1/2$ and to form a fermion. In view of isotopically selective laser cooling experiments with iron, it is of primary importance to accurately know the isotopic effects on both the Fe~I 372- and 358-nm transitions. The case of the 372-nm transition was studied in details in Ref.~\cite{Krins09}. In this paper, we focus on the 358-nm transition about which nothing is currently known. We report the measurement of the isotope shifts between all four stable iron isotopes, as well as the related hyperfine structure for $^{57}$Fe. The experimental setup differs significantly from that used for the study of the 372-nm transition~\cite{Krins09}. If the latter was based on a saturated-absorption spectroscopy system using an Fe-Ar hollow cathode, this scheme is hardly conceivable for the 358-nm transition since the energy of the lower state prevents it from being sufficiently populated in a hollow cathode operating continuously so as to generate detectable saturated-absorption signals. We used instead a laser-induced fluorescence scheme on an iron atomic beam produced in a high temperature oven filled with both natural and enriched samples of iron.

The paper is organized as follows. In Sec.~II, our experimental setup is described. We then expose our results and their analysis in Sec.~III. We finally draw conclusions in Sec.~IV.

\section{Experimental setup}

\begin{figure}
	\centering
	\includegraphics[width=7cm] {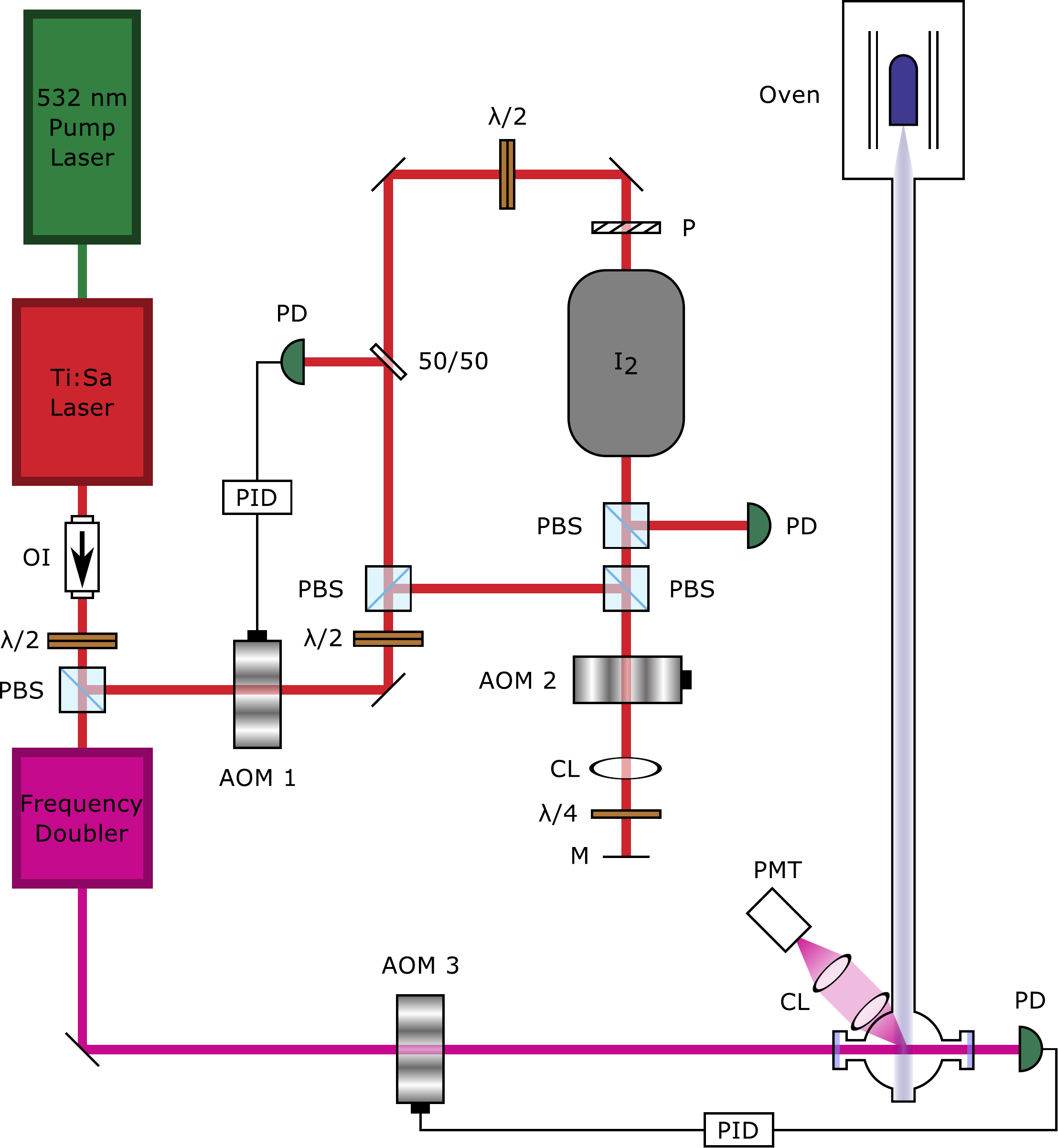}
	\caption{(Color online) Experimental arrangement used for the laser-induced fluorescence spectroscopy of the iron 358-nm transition (AOM: acousto-optical modulator, $\lambda/2$: half-wave plate, $\lambda/4$: quarter-wave plate, $50/50$: 50\% transmission plate beamsplitter, CL: converging lens, M: mirror, P: polarizer, PBS: polarizing beamsplitter, OI: optical isolator, PD: photodiode, PID: proportional-integral-derivative controller, PMT: Photomultiplier tube, I$_2$: molecular iodine cell).\label{setup}}
\end{figure}

The experimental setup is shown in Fig.~\ref{setup}. The radiation at 358~nm was produced by a frequency doubled Coherent$^\circledR$ Ti:Sapphire (Ti:Sa) laser set to 716~nm and pumped with a Coherent$^\circledR$ compact solid-state diode-pumped frequency-doubled Nd:Vanadate (Nd:YVO4) laser at 532 nm. The linewidth of the Ti:Sa laser was specified to be less than 1~MHz.
The laser radiations at 716~nm and at 358~nm were sent simultaneously to a saturation-spectroscopy setup on a molecular iodine cell and to a laser-induced fluorescence setup on an iron atomic beam, respectively. In this way, synchronized spectra of both molecular iodine and atomic iron at frequencies in a strict relation of 2 (within some fixed frequency shift of only a few hundred MHz induced by the presence of acousto-optical modulators (AOM) in the setup) could be recorded by scanning the Ti:Sa frequency. The molecular iodine spectra served as a calibration for the iron spectra. This was made possible since molecular iodine has precisely a 15-hyperfine-component line at almost exactly half the frequency of the analyzed iron line, namely the transition $\textrm{$B$ $^3\Pi^+_{0u} \leftarrow X$ $^1\Sigma^+_g$ $R(90) 3-10$}$ at 13957.8542(50)~cm$^{-1}$~\cite{Ger82,Ger85}, whereas half the wavenumber of the 358-nm iron line reads 13957.8445(11)~cm$^{-1}$~\cite{Nave94,NIST}. The hyperfine structure of the $R(90) 3-10$ iodine transition was studied in details in Ref.~\cite{iodinepaper}. A description of the saturation spectroscopy setup can be found in this reference and will not be repeated here. The fraction of the laser radiation at 716~nm sent to the molecular iodine saturation spectroscopy setup amounted typically to 100~mW.

The laser-induced fluorescence setup consisted of an iron atomic beam produced in a high temperature oven and directed through a 16-mm wide tube to an observation chamber where the 358-nm laser radiation crossed at right angle the atomic beam. The oven was operated at temperatures ranging from 1920~K to 1970~K to ensure a sufficient atomic flux. It yielded a statistical population of the lower state of the studied 358-nm iron line of about 0.35~\%. The center of the observation chamber was located at $\sim 1.30$~m of the oven aperture (2 mm in diameter), thereby limiting the atomic beam half divergence angle to a value not exceeding 0.4$^{\circ}$. To increase the signal-to-noise ratio, the laser beam power in the observation chamber was stabilized to about $150$~$\mu$W using a proportional-integral-derivative (PID) controller regulating the acoustic wave intensity in an acousto-optical modulator (AOM 3 in Fig.~\ref{setup}) so as to keep constant the power of the first order diffracted beam, as measured by a photodiode located at the exit of the observation chamber. The servo-loop had a time constant of about 20~$\mu$s. The laser beam section in the observation chamber was $\sim 1$ mm$^2$. The laser-induced fluorescence light emitted by the atoms was observed perpendicularly to both the laser and atomic beams through a large numerical aperture $f/1.5$ (0.33~sr) imaging system. The light was collected to a photomultiplier tube (PMT) covered with a 10-nm bandwidth interference filter to limit the background light at its input. The signal of the PMT was averaged by a boxcar averager.

\section{Results and discussion}

\subsection{Spectra}

\begin{figure}
	\centering
    \includegraphics[width=8cm] {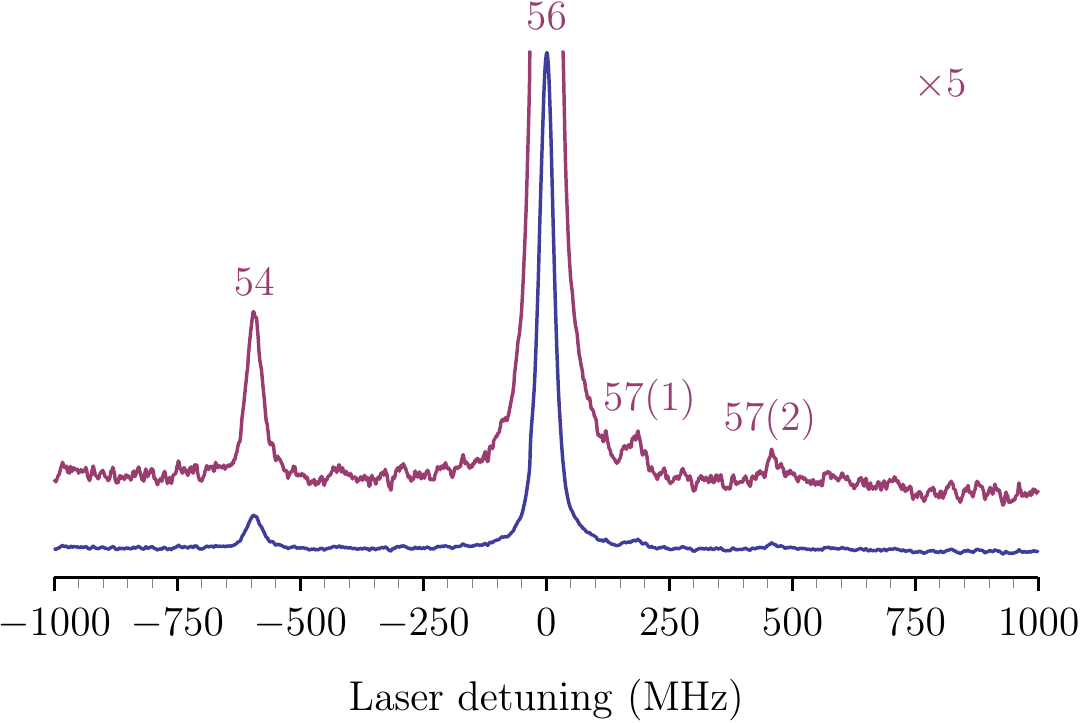}
	\caption{(Color online) Typical laser-induced fluorescence spectrum on a natural iron atomic beam at two different scales. The lines correspond to the $3d^74s \,\, a \, {}^5\!F_5 - 3d^74p \,\, z \, {}^5\!G^o_6$ Fe~I transition at 358~nm for the iron isotopes $^{54}$Fe, $^{56}$Fe and $^{57}$Fe. The natural abundance of the stable isotope $^{58}$Fe is so low (see text) that its contribution is not visible on the spectrum.\label{fernat}}
\end{figure}

A typical laser-induced fluorescence spectrum on the iron atomic beam is shown in Fig.~\ref{fernat}. For this spectrum, the oven was heated at 1970~K and filled with a natural iron powder (5.8\% of $^{54}$Fe, 91.8\% of $^{56}$Fe, 2.1\% of $^{57}$Fe, and 0.3\% of $^{58}$Fe~\cite{Ros98}). The strong central peak corresponds to the 358-nm transition for the most abundant isotope $^{56}$Fe. The lower frequency peak is the same transition for the isotope $^{54}$Fe. The two little peaks towards the higher frequencies are two hyperfine components of the $^{57}$Fe transition. Because of its very low abundance, the contribution of the isotope $^{58}$Fe is not visible on the spectrum. The fluorescence peaks had a linewidth of about 36~MHz. This value comes from the natural linewidth (16.20(45)~MHz~\cite{Bla79}), power broadened up to about 19~MHz, and from the atomic beam divergence broadening effect.
\begin{figure}
	\centering
	\includegraphics[width=7cm] {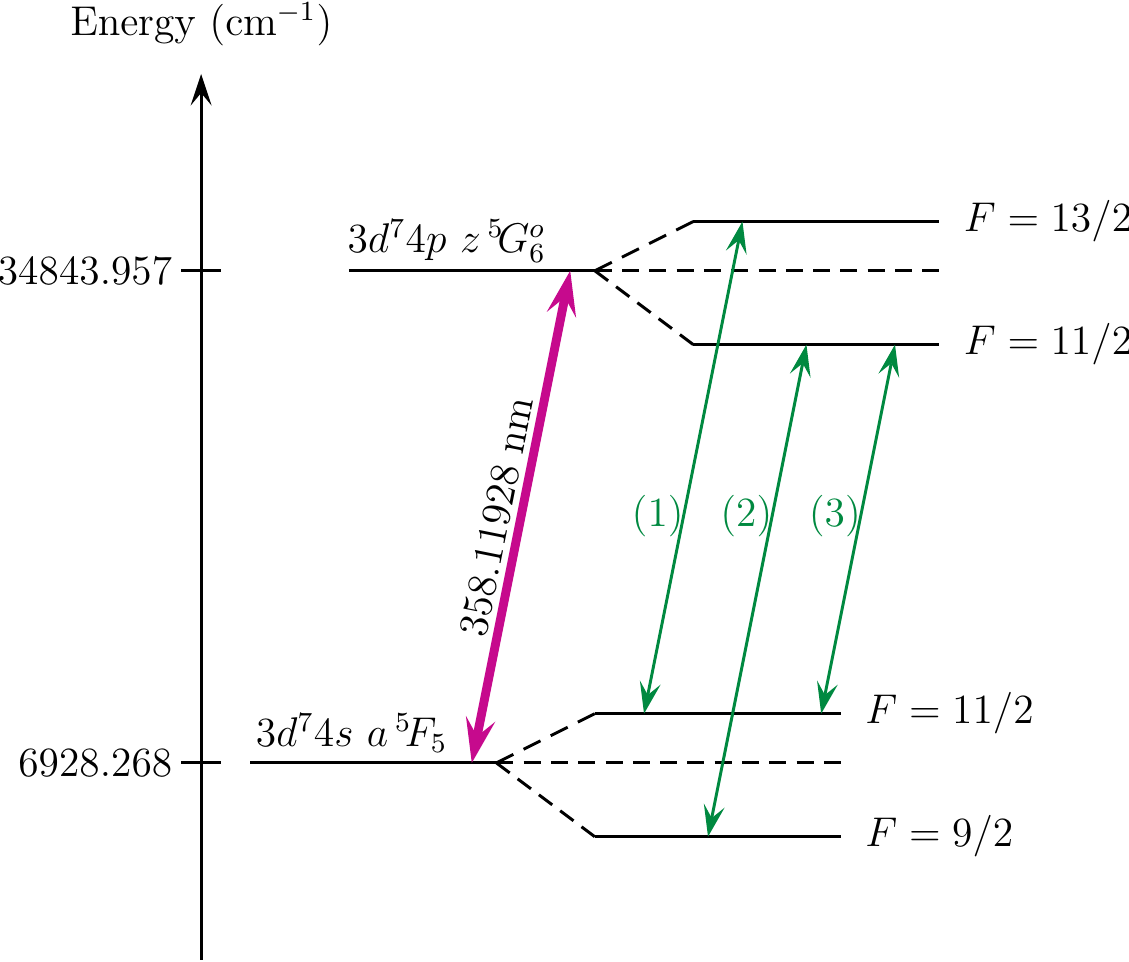}
	\caption{(Color online) Hyperfine structure components (1), (2) and (3) of the $3d^74s \,\, a \, {}^5\!F_5 - 3d^74p \,\, z \, {}^5\!G^o_6$ Fe~I 358-nm line for the isotope ${}^{57}$Fe (nuclear spin $I = 1/2$). The quoted state energies and air wavelength of the transition are the accurate values of Refs.~\cite{Nave94,NIST}.\label{Hfstruct}}
\end{figure}

The hyperfine structure of the $^{57}$Fe transition is illustrated in Fig.~\ref{Hfstruct}. The lower [upper] level is split into two hyperfine levels $F = 9/2$ and $F = 11/2$ [$F = 11/2$ and $F = 13/2$] that are shifted with respect to their unperturbed fine structure level by the amount (at the first-order perturbation theory) $A C/2$ with $A$ the hfs magnetic dipole coupling constant of the unperturbed level and
\begin{equation}
\label{C}
C = F(F+1) - I(I+1) - J(J+1),
\end{equation}
with $F$, $I$ and $J$ the total angular momentum, nuclear spin and total electronic angular momentum quantum numbers, respectively. Since the nuclear spin is $1/2$, no electric quadrupole effect is present. Three electric dipole transitions are associated to this hyperfine structure, as shown in Fig.~\ref{Hfstruct} with the labels $(1)$, $(2)$ and $(3)$. The relative theoretical intensities of these three hyperfine transitions are $100: 84.4: 1.3$, respectively~\cite{57intens}. The extreme weakness of the third transition explains why only two peaks are visible in the fluorescence spectrum of Fig.~\ref{fernat}.

\begin{figure}
	\centering
	\includegraphics[width=8cm] {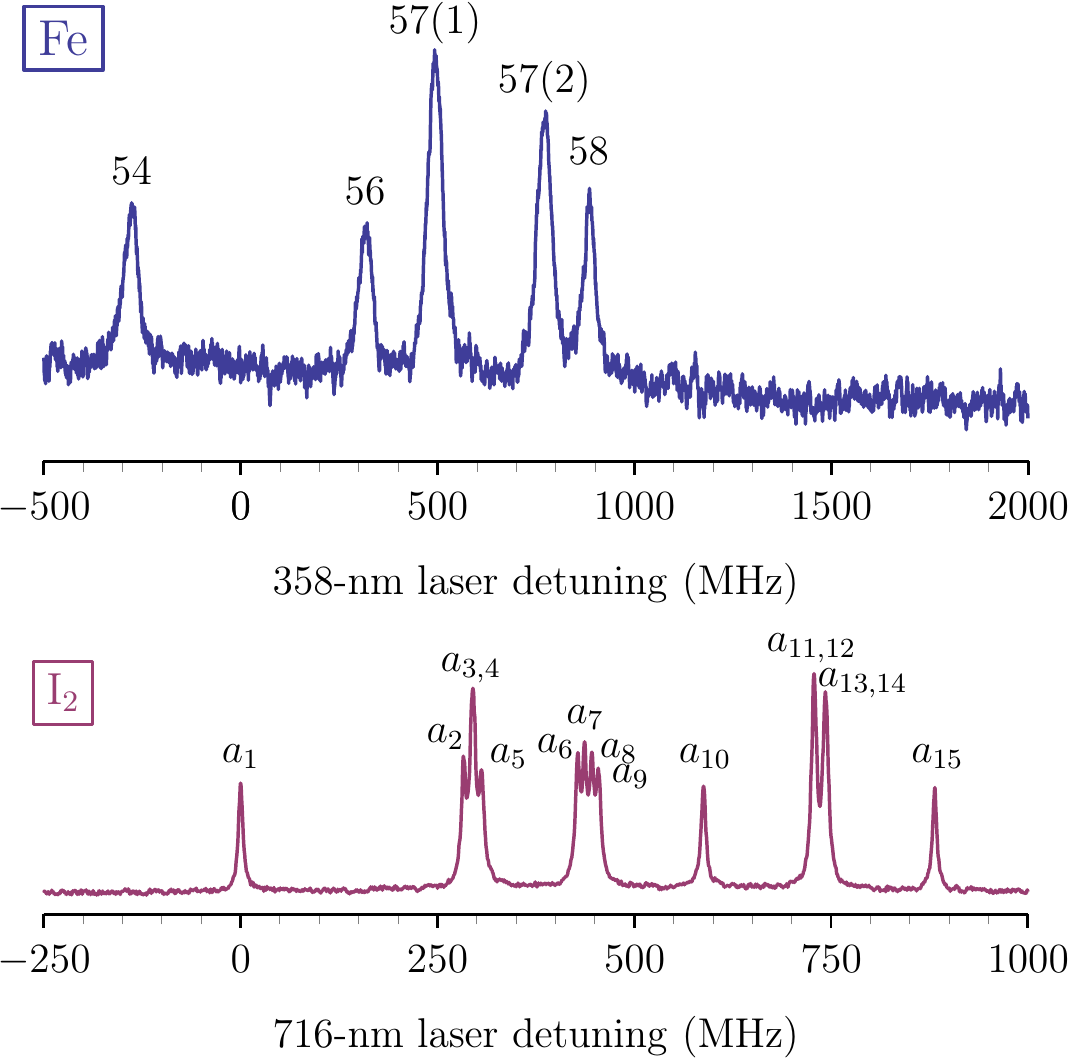}
	\caption{(Color online) Top: Same fluorescence spectrum as in Fig.~\ref{fernat}, recorded with an isotope-enriched powder of iron. Bottom: Doppler-free laser saturated-absorption spectrum of the 716-nm $R(90)3-10$ molecular iodine transition recorded synchronously with the iron spectrum and used for calibration purposes (see Ref.~\cite{iodinepaper} for the splitting values of all hyperfine components $a_i$ ($i=1-15$) of the transition). The laser detunings are here referenced with respect to the molecular iodine $a_1$ hyperfine component.\label{fluospec}}
\end{figure}

In order to enhance the contributions of the poorly abundant isotopes, the oven was filled with a home-made isotope-enriched iron powder composed of 14.3\% of $^{54}$Fe, 14.3\% of $^{56}$Fe, 57.2\% of $^{57}$Fe and 14.3\% of $^{58}$Fe. A typical laser-induced fluorescence spectrum recorded with this powder is shown in Fig.~\ref{fluospec}, along with the saturated-absorption spectrum of molecular iodine recorded synchronously and used for calibration purposes. Here, the intensity of the $^{58}$Fe contribution is approximately equal to the $^{54}$Fe and $^{56}$Fe lines', while the two peaks related to the $^{57}$Fe hyperfine structure are enhanced in accordance with the chosen abundance in the home-made iron powder. The intensity of the 57(2) peak is approximately equal to 75\% of the 57(1) peak's. It is a bit lower than the expected 84.4\% theoretical intensity, but the difference is easily explained from an optical pumping phenomenon between the $F=9/2$ and $F=11/2$ hyperfine states of the $3d^74s \,\, a \, {}^5\!F_5$ lower energy level. When passing through the laser beam tuned to resonance with the 57(2) hyperfine transition, some atoms of the atomic beam get lost from the fluorescence process when they decay through the very weak 57(3) hyperfine transition and accumulate little by little in the $3d^74s \,\, a \, {}^5\!F_5$ $(F=11/2)$ state. The 57(1) hyperfine transition is immune from such a phenomenon. The third hyperfine component remains itself invisible in the spectrum of Fig.~\ref{fluospec}, in agreement with its expected intensity lower than the signal-to-noise ratio.

\subsection{Isotope shifts and hyperfine structure constants}

From the analysis of 33 laser-induced fluorescence iron spectra similar to that of Fig.~\ref{fluospec}, we deduced the isotope shifts $\delta\nu_{56,54}$, $\delta\nu_{57,56}$ and $\delta\nu_{58,56}$ of the 358-nm Fe~I transition, as well as the hfs magnetic dipole coupling constant $A'$ of the upper level of the transition. These values are summarized in Table~\ref{IS_A}. In this table, the uncertainties quoted represent the statistical errors (one standard deviation of the mean of the sample of the 33 spectra). All values presented in this table are determined for the first time. For the isotopes $54$ and $58$ with no nuclear spin, the isotope shifts $\delta\nu_{56,54}$ and $\delta\nu_{58,56}$ are directly obtained from the frequency shifts of their respective transitions with respect to the $^{56}$Fe line. In the case of the isotope 57, the situation is a bit more complicated because of the intertwined hyperfine structure effect. The isotope shift $\delta\nu_{57,56}$ is the frequency shift that would be observed between isotopes 57 and 56 without the hyperfine structure. In its presence, the frequency shift of each hyperfine component~$(i)$ ($i=1-3$) reads
\begin{equation}\label{IS57}
\delta\nu_{57(i),56}=\delta\nu_{57,56}+\frac{1}{2}A'C'_{(i)}-\frac{1}{2}AC_{(i)}
\end{equation}
with $A$ [$A'$] the hyperfine constant of the lower [upper] level and $C_{(i)}$ [$C'_{(i)}$] the constant of Eq.~\eqref{C} for the lower [upper] level of the hyperfine component~$(i)$. From the observed frequency shifts $\delta\nu_{57(1),56} = 175.33(55)$~MHz and $\delta\nu_{57(2),56} = 452.11(63)$~MHz, as well as the known value of the hyperfine constant $A$ of the lower state ($A=87.246(3)$~MHz~\cite{ahfs}), Eq.~(\ref{IS57}) yields a set of 2 equations for the 2 sought unknowns $\delta\nu_{57,56}$ and $A'$.

\begin{table}
\caption{Isotope shifts between the given iron isotopes for the Fe~I 358-nm transition, as well as the hfs magnetic dipole coupling constant $A'$ of the upper level of the transition.}
    \label{IS_A}
    \renewcommand{\arraystretch}{1.4}
\begin{center}
    \begin{tabular}{cccccc|ccc}
    \hline\hline
    $\underset{\textrm{\small (MHz)}}{\delta \nu_{56,54}}$
 & & $\underset{\textrm{\small (MHz)}}{\delta \nu_{57,56}}$ & & $\underset{\textrm{\small (MHz)}}{\delta \nu_{58,56}}$ & & & & $\underset{\textrm{\small (MHz)}}{A'_{~}}$\\
     \hline
$592.28(61)$ & & $299.72(57)$ & & $566.02(47)$ & & & & $31.241(48)$\\
\hline \hline
    \end{tabular}
\end{center}
    \renewcommand{\arraystretch}{1}
\end{table}

\subsection{Field and specific mass shifts}

\begin{figure}
	\centering
    \includegraphics[width=8cm] {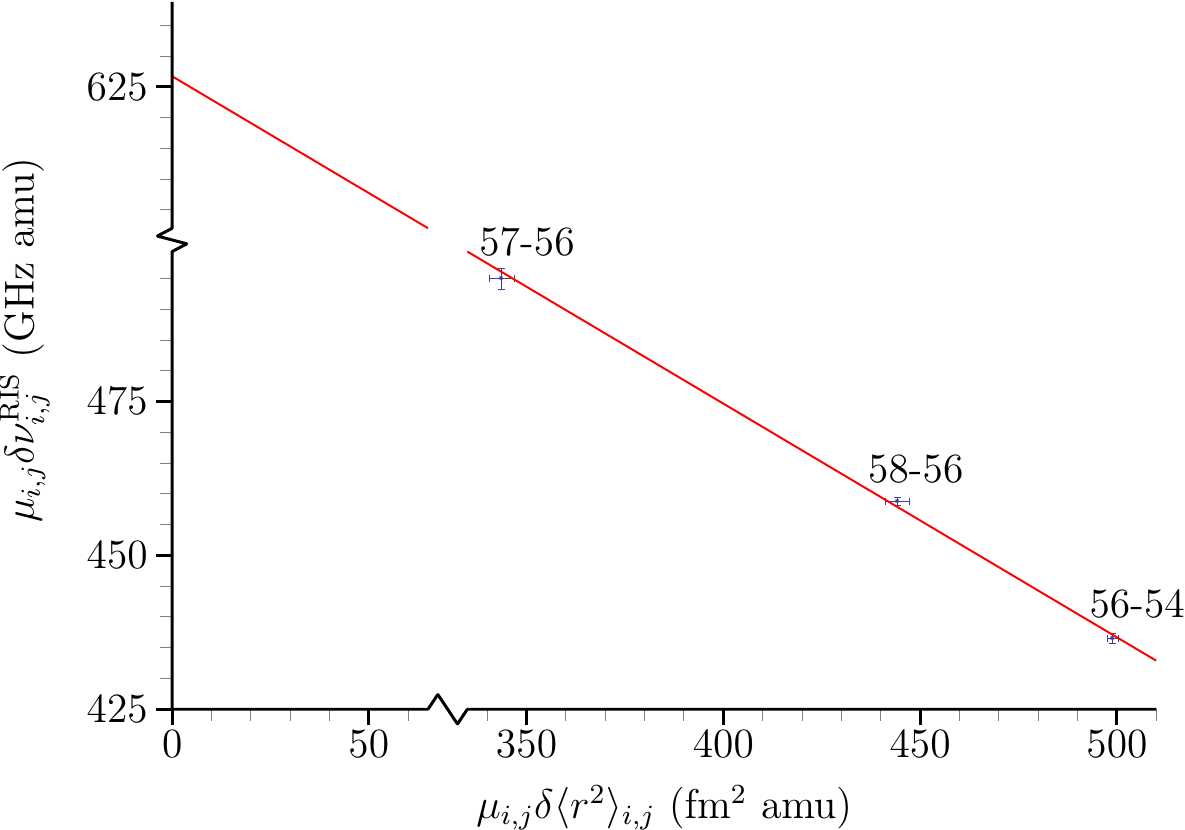}
	\caption{(Color online) King plot of the Fe~I 358-nm transition. The error bars stand for one standard deviation. The vertical axis intercept and slope of the linear regression line yield the specific mass shift $k_\mathrm{SMS}$ and field shift $F$ coefficients, respectively.\label{KPfig}}
\end{figure}

The isotope shift $\delta \nu_{i,j}$ between two isotopes $i$ and $j$ is known to be the sum of two contributions~\cite{King}~:
\begin{equation}
\delta \nu_{i,j}=k\mu^{-1}_{i,j}+F\delta\langle r^2\rangle_{i,j},
\end{equation}
where $k$ and $F$ are the mass and field shift coefficients of the transition, respectively, $\delta \langle r^2 \rangle_{i,j}$ is the difference in mean square nuclear charge radii between the two isotopes, and $\mu_{i,j} = m_{i} m_{j}/(m_{i}-m_{j})$ with $m_{i}$ and $m_{j}$ the masses of the 2 isotopes. The mass shift coefficient $k$ is itself the sum of two contributions, the normal mass shift coefficient $k_\mathrm{NMS}$, taking into account the electron reduced mass correction, and the specific mass shift coefficient $k_\mathrm{SMS}$, originating
from the change in the correlated motion of all the electrons~: $k=k_\mathrm{NMS}+k_\mathrm{SMS}$. While the normal mass shift coefficient merely reads $k_\mathrm{NMS}=m_e\nu/u$, where $m_e$ is the electron mass, $\nu$ is the transition frequency and $u$ is the atomic mass unit, the specific mass shift coefficient $k_\mathrm{SMS}$ is much more complicated to evaluate \textit{ab initio}. By subtracting the normal mass shift $k_\mathrm{NMS}\mu^{-1}_{i,j}$ from the isotope shift $\delta \nu_{i,j}$, one gets the so-called residual isotope shift $\delta\nu^\mathrm{RIS}_{i,j}$, checking
\begin{equation}
\mu_{i,j}\delta\nu^\mathrm{RIS}_{i,j}=k_\mathrm{SMS}+F\mu_{i,j}\delta\langle r^2\rangle_{i,j}.
\end{equation}
The plot of $\mu_{i,j}\delta\nu^\mathrm{RIS}_{i,j}$ as a function of $\mu_{i,j}\delta\langle r^2\rangle_{i,j}$ for several isotope pairs $(i,j)$ (the so-called King plot~\cite{King}) yields a straight line of slope $F$ and vertical axis intercept $k_\mathrm{SMS}$.

The King plot associated with our experimental isotope shifts is illustrated in Fig.~\ref{KPfig}. We considered for this plot $\delta\langle r^2\rangle_{56,54} = 0.330(1)$ fm$^2$, $\delta\langle r^2\rangle_{57,56} = 0.108(1)$~fm$^2$, and $\delta\langle r^2\rangle_{58,56} = 0.274(2)$~fm$^2$~\cite{radii}. The linear regression of the plotted points taking into account both the horizontal and vertical error bars~\cite{York2004} yielded
\begin{equation}
\begin{aligned}\label{Fk}
k_\mathrm{SMS} & = 626.7(7.1)\textrm{ GHz amu}, \\
F & = -0.380(15)\textrm{ GHz/fm}^2,\\
\end{aligned}
\end{equation}
where the quoted uncertainties represent the statistical error (one standard deviation). From these values, it is possible to extract the contribution of the specific mass shift $k_\mathrm{SMS}\mu^{-1}_{i,j}$ and the field shift $F \delta\langle r^2\rangle_{i,j}$ for any isotope pairs $(i,j)$. These contributions are summarized in Table~\ref{RIS}. As can be seen from this table, the normal mass shift terms account for half the isotope shifts and the field shifts contribute negatively.

\begin{table}
\caption{Isotope shift (IS), residual isotope shift (RIS), specific mass shift (SMS), and field shift (FS) between the given isotopes for the Fe~I 358-nm transition.}
    \label{RIS}
\renewcommand{\arraystretch}{1.4}
\begin{center}
    \begin{tabular}{c|cccc}
    \hline\hline
$\underset{\textrm{\small pair}}{\textrm{\small Isotope}_{~}}$ & $\underset{\textrm{\small(MHz)}}{\mathrm{IS}_{~}}$ & $\underset{\textrm{\small(MHz)}}{\mathrm{RIS}_{~}}$ & $\underset{\textrm{\small(MHz)}}{\mathrm{SMS}_{~}}$ & $\underset{\textrm{\small(MHz)}}{\mathrm{FS}_{~}}$ \\
     \hline
$56-54$ & $592.28(61)$ & $288.66(61)$ & $414.5(4.7)$ & $-125.4(4.9)$\\
$57-56$ & 299.72(57) & 155.49(57) & 196.9(2.2) & $-41.0(1.8)$\\
$58-56$ & 566.02(47) & 282.90(47) & 386.5(4.4) & $-104.1(4.3)$\\
\hline \hline
    \end{tabular}
\end{center}
    \renewcommand{\arraystretch}{1}
\end{table}

\section{Conclusion}

In this paper, we have reported at the third percent level the first experimental determination of the isotope shifts of the $3d^74s \,\, a \, {}^5\!F_5 - 3d^74p \,\, z \, {}^5\!G^o_6$ Fe~I line at 358~nm between all four stable isotopes ${}^{54}$Fe ($I=0$), ${}^{56}$Fe ($I=0$), ${}^{57}$Fe ($I=1/2$) and ${}^{58}$Fe ($I=0$), as well as the hyperfine structure of that line for ${}^{57}$Fe. The knowledge of these frequency shifts is of primary importance in the context of any laser cooling experiment for iron atoms since the Fe~I 358-nm line has been identified as the first accessible iron transition suitable for that purposes~\cite{iodinepaper}. A King plot analysis has further yielded the field and specific mass shift coefficients of the transition. The measurements were carried out by means of laser-induced fluorescence spectroscopy on an iron atomic beam produced from an oven at temperatures ranging between 1920 and 1970~K. The oven was filled with a home-made isotope-enriched powder to enhance the contribution of the isotopes poorly abundant in natural samples.

\acknowledgments
The authors acknowledge the financial support from the Belgian F.R.S.-FNRS through IISN Grant 4.4512.08. It is a pleasure to thank M.~Godefroid for helpful discussions.

\end{document}